\documentclass{article}
\usepackage{spconf,amsmath,graphicx}
\usepackage{multirow}
\usepackage{booktabs}
\usepackage{steinmetz}
\usepackage{amssymb}
\usepackage{xcolor}
\usepackage[flushleft]{threeparttable}
\usepackage{arydshln}

\title{Exploring End-to-End Multi-channel ASR with Bias Information for Meeting Transcription}
\name{Xiaofei Wang, Naoyuki Kanda, Yashesh Gaur, Zhuo Chen, Zhong Meng, Takuya Yoshioka}
\address{Microsoft Corp., Redmond, WA, USA}
\begin{document}
\ninept
\maketitle

\begin{abstract}
%
%
%
Joint optimization of multi-channel front-end and automatic speech recognition (ASR) has attracted much interest. While promising results have been reported for various tasks, past studies on its meeting transcription application were limited to small scale experiments. 
It is still unclear whether such a joint framework can be beneficial for a more practical setup where a massive amount of single channel training data can be leveraged for building a strong ASR back-end. In this work, we present our investigation on the joint modeling of a mask-based beamformer and Attention-Encoder-Decoder-based ASR in the setting where we have 75k hours of single-channel data and a relatively small amount of real multi-channel data for model training. We explore effective training procedures, including a comparison of simulated and real multi-channel training data. To guide the recognition towards a target speaker and deal with overlapped speech, we also explore various combinations of bias information, such as direction of arrivals and speaker profiles. We propose an effective location bias integration method called deep concatenation for the beamformer network. In our evaluation on various meeting recordings, we show that the proposed framework achieves a substantial word error rate reduction.

\end{abstract}

\section{Introduction}
\label{sec:intro}
 Meeting transcription with speaker annotation is one of the challenging tasks in automatic speech recognition (ASR) field \cite{xiong2016achieving,yoshioka2019advances,chen2020continuous}.
It is difficult not only because of its natural conversational content but also because of complicated acoustic conditions often with speaker overlaps \cite{li2017acoustic,yoshioka2018recognizing,yoshioka2018multi}.
To improve the model robustness in challenging acoustic conditions, multi-channel front-end speech processing is often introduced to separate target speaker signals
from the background noise, reverberation and interfering speakers~\cite{yoshioka2019advances}. 

One of the common approaches for ASR front-end processing is time-frequency mask-based beamforming by using a mask estimation network
\cite{wang2018supervised,kinoshita2017neural,heymann2016neural,erdogan2016improved,yu2017permutation,boeddeker2018front}.
It has shown promising results in 
various ASR benchmarks,
such as AMI \cite{carletta2005ami} and CHiME challenges \cite{barker2018fifth,watanabe2020chime,erdogan2016multi}.
However, the mask estimation network is normally trained by using simulation data in order to prepare an accurate reference mask, leading to potential mismatch problems for real data. 
Furthermore, the network is often trained by optimizing a signal-level criterion, which is not necessarily optimal for ASR.

Recent studies suggest that joint optimization of multi-channel front-end and ASR can yield
better recognition results than sequential processing scheme with separately optimized front-end and ASR modules~\cite{kanda2018hitachi,kanda2019acoustic,subramanian2019speech,subramanian2020far,von2020multi,zhang2020end}. 
With the joint optimization approach, 
all modules are connected in one computational graph and trained by back-propagating the ASR training loss to ensure optimal ASR performance (for the training set). 
The joint optimization also enables us to leverage transcribed real data for which reference clean signals are unavailable. 
This is advantageous for large scale training, as the signal level reference is usually extremely difficult to collect for real data compared with transcription. 
It would also help reduce the mismatch of training and testing conditions. 

%

While the joint optimization approach is promising, previous studies on meeting (or more broadly conversation)
transcription were either all conducted on simulated~\cite{sainath2017multichannel}, small-scale data~\cite{kanda2018hitachi,kanda2019acoustic, meng2017deep}, or did not particularly emphasize the capability to deal with overlapped speech~\cite{li2017acoustic,minhua2019frequency}.
This is because
acquiring accurate transcription for multi-channel meeting recordings is much more difficult 
than doing so for voice commands or other single-speaker dictation tasks.
On the other hand, it is relatively easy to collect single-channel audio as the ASR training data.
Therefore, it is practically important to explore the joint front-end and back-end modeling with a massive amount of single-channel training data and a limited amount of multi-channel training data.
However, to the best of our knowledge, there has still been no 
exploration in this direction, especially with a scale of 75k hours of training data as is done in this paper.


Besides the joint modeling of front-end and back-end, due to the long recording nature in meetings, it is feasible for meeting transcription systems to leverage long context information that would help extract a target speech from overlapped sentences \cite{yoshioka2019advances}.
For instance, the direction of arrival (DOA) information can be encoded as angle features to get time-frequency masks that are biased toward the target speaker direction to reconstruct the targer signal from the mixture \cite{chen2018multi}. The speaker embeddings could also be utilized to make the mask estimator biased to the target speaker \cite{subramanian2020far,delcroix2020improving,xiao2019single}. Compared with the blind separation approach, target speech recognition does not have the permutation problem and thus could potentially lead to better result.

In this work, we describe our investigation on the effectiveness of 
an end-to-end multi-channel ASR (E2E MCASR) for meeting transcription under the condition 
that we leverage a massive amount of single-channel data for the back-end training while the multi-channel training data is still limited. The front-end and back-end models are then jointly optimized, which takes a multi-channel signal as input and outputs the word sequence of a target speech. 
We firstly review several key techniques, including mask-based beamforming, the usage of multi-channel features, Attention-Encoder-Decoder (AED)-based ASR, and joint training of these models. 
We then introduce bias information in the front-end module to further enhance the joint models for meeting transcription. 
We also explore a session-based semi-blind decoding strategy in which the bias information can be accurately estimated, where the SSL and speaker profile extraction can be simply replaced by online approaches. 
In our evaluation on various meeting recordings, we show that the proposed framework achieves substantial word error rate reduction.

\begin{figure*}[htb]
\begin{centering}
\includegraphics[scale=0.7]{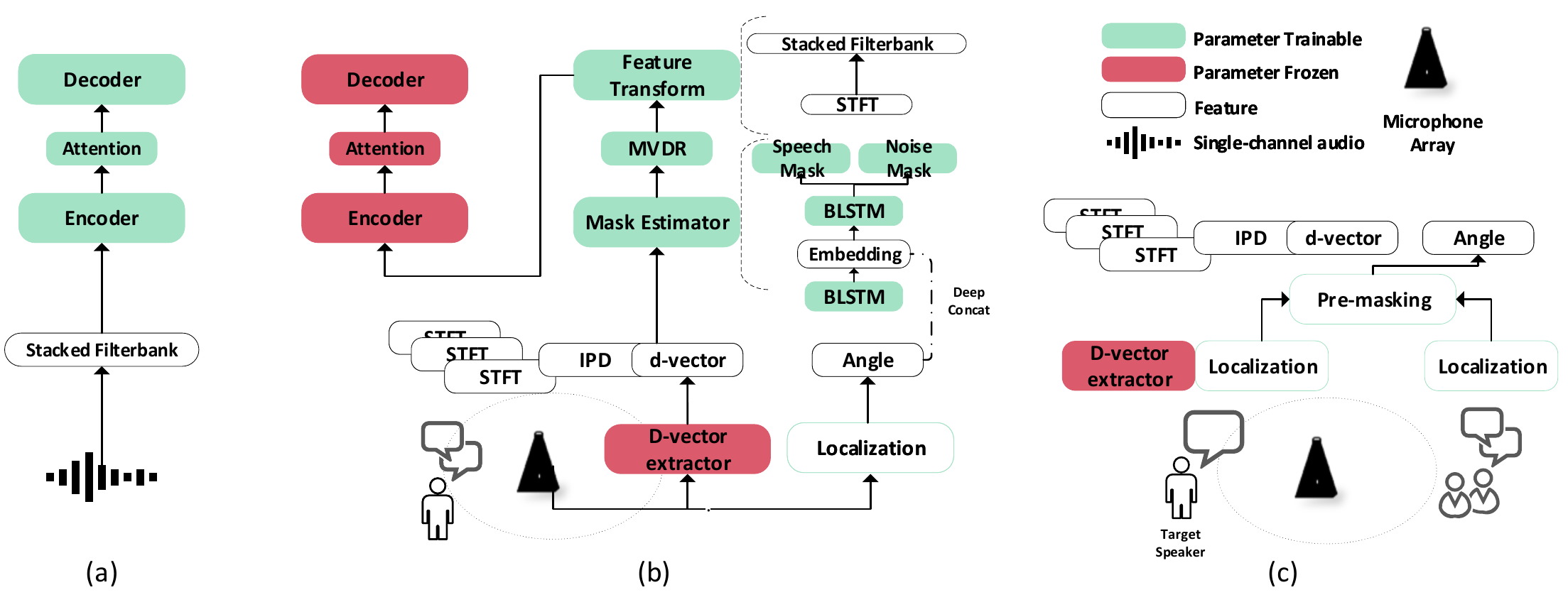}
\caption{The E2E MCASR framework for meeting transcription. (a) The AED-based ASR trained with single-channel data; (b) Joint front-end/back-end model trained with multi-channel meeting data and bias information, where the details of each module are presented within the dotted lines; (c) Session-based decoding strategy for target-speaker ASR.}
\label{fig:dftnet}
\end{centering}
\vspace*{-4mm}
\end{figure*}

\section{E2E MCASR framework}
Figure \ref{fig:dftnet} shows how we leverage the E2E MCASR framework for the meeting transcription task. The framework consists of several modules.

\subsection{Front-end module}
\subsubsection{Neural Beamforming}
The neural beamformer is defined with parameter $\theta _{\text{bf}}$ as follows:
\begin{equation}
 O = {\rm Beamformer}(X; \theta _{\text{bf}}),
 \label{eq:fcsnet}
\end{equation}
where $O$ and $X$ denote the beamformed and noisy short-time Fourier transform (STFT) of signals for all frames and frequency bins, respectively.
For beamforming, which converts the multi-channel signals $\mathbf{x}(t, f)\in X$ ($t, f$ are time and frequency bin indexes, respectively) to single-channel beamformed signal $o(t,f)\in O$
\begin{equation}
o(t,f) = \mathbf{w}_{\textrm{MVDR}}^\textrm{H} (f) \mathbf{x}(t, f),
\label{apply_bf}
\end{equation}
time-invariant minimum variance distortionless response (MVDR) beamformer $\mathbf{w}_{\textrm{MVDR}}(f)$ in the frequency domain has been employed. Mathematically, the MVDR filter can be calculated with the following formula
\begin{equation}
\label{eqn:mvdr}
\mathbf{w}_{\textrm{MVDR}} (f) = \frac{\mathbf{\Phi}_{\text{N}} (f)^{-1}\mathbf{\Phi}_{\text{S}} (f)}{\text{Tr}(\mathbf{\Phi}_{\text{N}} (f)^{-1}\mathbf{\Phi}_{\text{S}} (f))} \mathbf{u},
\end{equation}
where $\mathbf{u} \in \{0, 1\} ^{M}$ is a one-hot vector to choose a reference microphone \cite{ochiai2017unified}. 
$\text{Tr}(\cdot)$ denotes the trace operation.
${\mathbf{\Phi}_\text{N}}(f)$ and ${\mathbf{\Phi}_\text{S}}(f)$ denote the power spectral density (PSD) matrices of noise and speech at frequency bin $f$, respectively, which are calculated by
\begin{equation}
\label{eqn:psd}
\mathbf{\Phi}_{v} (f) = \sum\limits_{t=1}^T M_v(t,f)\mathbf{d}(t, f)\mathbf{d}^\textrm{H}(t, f) \  \textrm{where} \  v \in \{\text{S}, \text{N}\}
\end{equation}
Speech mask $M_{\text{S}}(t,f) \in [0, 1]$ and noise mask $M_{\text{N}}(t,f)\in [0, 1]$ are estimated by applying a bidirectional long short-term memory (BLSTM)-based network to each microphone. The masks are then averaged over the microphone channels for PSDs estimation. Note that a two-head (speech head and noise head) training scheme has been employed \cite{kinoshita2017neural}\cite{heymann2016neural}, where both speech masks and noise masks are estimated. 
The respective masks indicate the time-frequency bins that are dominated by speech and noise.
The advantage of the two-head training scheme is that the BLSTM network has been given additional hints for better learning the relationship between the speech mask and noise mask, potentially leading to better mask predictions. 

\subsubsection{Unbiased multi-channel features}
Multi-channel features have been shown effective for the front-end network. 
In \cite{subramanian2019speech,von2020multi,zhang2020end}, the spectrum of $M$-channel signals were fed into the mask estimation network (referred to as STFT features in this work). In \cite{yoshioka2018multi,subramanian2020far}, the authors utilized inter-microphone phase difference (IPD) features to capture the spatial pattern from a multi-channel input, and observed a significant performance boost in both speech separation and recognition tasks. 
The IPD features can be calculated by 
\begin{equation}
\label{eqn:ipd}
\mathbf{IPD}_{i}(t,f) = \angle \frac{x_{i}(t,f)}{x_{1}(t,f)}, i=2,...,M,
\end{equation}
where $\angle$ outputs angle of the input argument, $M$ denotes the number of the microphones, $i$ is the microphone index. Utterance-wise normalization \cite{chen2018multi} is applied before the IPD features are concatenated with the STFT features. Unlike the work of \cite{subramanian2020far}, we concatenate the IPD features calculated with respect to the first microphone to the STFT of each channel, which are fed to the mask estimation network. This results in multiple mask estimates, which are then averaged across the channels.

\subsection{Joint front-end and ASR via feature transformation layer}

In the joint front-end and back-end training scheme, the signal is processed by the front-end and back-end layers sequentially, where a feature transformation layer is applied that maps the beamformed output from front-end layers to log mel filterbanks. 
Considering that back-end ASR usually takes multiple contextual frames as the input, the frame-based STFT feature is stacked in the feature transformation layer as follows,
\begin{equation}
O_P = {\rm Frame2Superframe}({\rm LogMel(|O|)}, P),
\end{equation}
where $P$ is the number of frames to be stacked. Finally, global mean-variance normalization (GMVN) is applied on top of the stacked filterbank features $O_P$ to derive the input $\hat O=\{o_1,...,o_T\}$ for back-end ASR, which is
\begin{equation}
\hat O = {\rm GMVN}(O_P),
\end{equation}
where the mean and variance vectors are obtained from the single-channel data for back-end training rather than the multi-channel training data. 

We select to combine the front-end and back-end via the feature transformation layer in order to fully utilize both single-channel and multi-channel training data.
The details of the training scheme will be explained in Section \ref{sec:twostage}.


\subsection{AED-based ASR}
The AED-based ASR consists of encoder, attention, and decoder modules. 
The AED model follows three steps to generate text sequence $Y=\{y_1, ..., y_n, ...\}$ given acoustic feature $\hat O=\{o_1,...,o_T\}$.
Firstly, the encoder module converts the input sequence $\hat O$ into a sequence of embeddings $H^{enc}$,
\begin{align}
H^{enc} &=\{h^{enc}_1,...,h^{enc}_T\}={\rm Encoder}(\hat O).  \label{eq:enc} 
\end{align}
Then, for every decoder step $n$, the attention module outputs context vector $c_n$ with attention weight $\alpha_n$
given decoder state vector $q_n$, the previous attention weight $\alpha_{n-1}$, and $H^{enc}$ as follows,
\begin{align}
c_n, \alpha_n &= {\rm Attention}(q_n, \alpha_{n-1}, H^{enc}). \label{eq:att} 
\end{align}
Finally, 
the output distribution $y_n$ 
is estimated given
the context vector $c_n$
and the decoder state vector $q_n$ as follows,
\begin{align}
 q_n &={\rm DecoderRNN}(y_{n-1}, c_{n-1}, q_{n-1}), \\
 y_n&= {\rm DecoderOut}(c_n,q_n). \label{eq:dec}
\end{align}

The ${\rm DecoderRNN}$ consists of several RNN layers and 
an affine transform followed by a softmax output layer are used for ${\rm DecoderOut}$.
The model is trained to minimize the cross entropy loss between $Y$ and reference label $R=\{r_1,...,r_N, r_{N+1}=\langle eos\rangle\}$ as follows
\begin{align}
\mathcal{L}^{CE}=\sum_{n=1}^{N+1}{\rm CE}(y_n, r_n), 
\end{align}
where ${\rm CE()}$ denotes the cross entropy function, $N$ is the number of symbols in the reference $R$, and $\langle eos\rangle$ is the special symbol indicating the end of a sentence.

\section{E2E MCASR with Target Bias Information}
\label{sec:biased}
The bias information, such as locations of attendees in the meeting~\cite{chen2018multi} and attendees' profiles \cite{subramanian2020far,delcroix2019end}, guides the framework towards target speech recognition, and is shown effective in both separation and recognition task. We incorporate the additional information for each target speaker to further improve the E2E MCASR system in this work. In meeting scenario, both location and speaker bias can be formed by averaging statistics from long recording with the help of diarization algorithms. For example, from the diarization system~\cite{yoshioka2019advances}, we can robustly identify the active region for each speaker, from which we can roughly estimate the location and speaker embedding by the sound source localization and speaker identification systems, under the assumption that the speakers don't move frequently during the meeting.

\subsection{Location bias in model training}

\subsubsection{Angle feature}
We use angle feature \cite{chen2018multi} as spatial bias in this work. The angle feature is calculated by the cosine distance between the steering vector that derived from DOA \cite{benesty2008microphone} and the complex spectrum of each channel that is normalized by the first microphone or the IPD on all selected microphone pairs, as suggested in Eqn.(\ref{eqn:angle}). 

\begin{equation}
\label{eqn:angle}
\mathbf{A}(t,f) = \frac{e_{i}(f)\frac{x_{i}(t,f)}{x_{1}(t,f)}}{|e_{i}(f)\frac{x_{i}(t,f)}{x_{1}(t,f)}|}, i=2,...,M,
\end{equation}
where $e_{i}(f)$ is the steering vector for target speaker at microphone i and frequency f.

The target DOA for angle feature extraction is estimated by using the maximum likelihood sound source localization algorithm \cite{zhang2007maximum} with a $3^\circ$ resolution.

During testing, following \cite{chen2018multi}, an additional pre-masking step is applied in the decoding stage to increase the discrimination resolution between the target speaker and the others in the same meeting session. 
This step has one more assumption that the DOAs from all the other attendees are known. 
In this work, we use the session-based DOA estimation for each speaker. Then the pre-masking is applied to angle feature which is obtained using the target DOA. The new angle feature $\hat{\mathbf{A}}_{t,f}$ used for decoding is
\begin{equation}
\label{eqn:premasking}
\hat{\mathbf{A}}(t,f) = \mathbf{A}(t,f)*\mathbf{Relu}(\mathbf{Sign}(\mathbf{A}(t,f)-\mathbf{A}^n(t,f))),
\end{equation}
where the $\mathbf{Relu}(\mathbf{Sign}(\bullet))$ function outputs 0 if the input is negative and 1 otherwise. $\mathbf{A^n}$ is angle features derived from the other speakers with different locations. Eqn.(\ref{eqn:premasking}) means the time-frequency bins dominated by other speakers are set to zero, motivated by the sparsity property of speech spectrogram.
Here, we restrict $|DOA(\mathbf{A})-DOA(\mathbf{A^n})|>\theta$, which means we only remove the bins that are dominated by the speakers whose directions with respect to the target speaker are larger than angle $\theta$. Empirically, we set it to $30^\circ$ (overall $360^\circ$) in the meeting transcription experiments.

\subsection{Speaker bias in model training}

The speaker information is also used as a biased feature for E2E MCASR in this work. 
However, compared to location information that all the directions can be covered in the training set, the speakers in testing phase may be unseen, e.g. the attendee does not have profiles or was not invited to the meeting. 

A d-vector extractor trained on 
VoxCeleb Corpus \cite{nagrani2017voxceleb,chung2018voxceleb2}
is employed in this work \cite{zhou2019cnn}. A 128-dim d-vector is extracted from the first channel of the multi-channel audios and concatenated with each channel's input features of the mask estimation network, illustrated by Fig.\ref{fig:dftnet}(b). The d-vector extractor is a universal auxiliary network that provides additional speaker information for learning the masks of target speech. The reason that we use d-vector is because accurate speaker profiles or speaker anchors~\cite{subramanian2020far,delcroix2020improving} are not always available in the real meetings.

\subsection{Deep concatenation}
Indicated by Fig.\ref{fig:dftnet}(b), we can concatenate the input of the mask estimation network with the same angle feature/d-vector, where the operation is similar to the IPD feature. Alternatively, the bias feature can be concatenated to the embedding, which is the output of the first layer of the mask estimation network. We name it deep concatenation in this work and refer to regular feature concatenation as shallow concatenation.

It is inspired by the SpeakerBeam work that allows tracking speech from a target speaker using a speaker adaptation layer \cite{delcroix2020improving,delcroix2019end}. Simply using the biased feature at the input level realizes the bias at the input layer, while the input layer bias can be eliminated because the input layer also has to encode the other features, e.g. unbiased IPD feature, which is insufficient to guide the network to focus on the target speech. This operation is simple yet effective for target speech recognition, especially in the overlapped regions.

\subsection{Session-based bias estimation}
Transcribing the speeches of the target speaker can be successfully achieved by knowing the DOA and d-vector in advance. Due to that, the joint model is designed for offline meeting transcription, session-based bias information is valuable in this case. Given the ground-truth diarization, we use the longest non-overlapped segment in a specific meeting session to extract the DOA and d-vector for each speaker, then apply this bias information to all the segments in the same session. An implicit assumption is that all the speakers in the session are not moving frequently, which is usually the case in real meetings. The reason we would like to use this information is to simplify the procedures in decoding and evaluate how the E2E MCASR performs given the almost accurate bias information. In a meeting scenario, both the DOA and d-vector of the target speaker can be tracked online via diarization or auxiliary cameras \cite{yoshioka2019advances}.


\section{Meeting Transcription Experiment}

\subsection{Single-channel training data for AED-based back-end}

The training set for the AED-based model is 75 thousand (K) hours of single-channel transcribed data, recorded in various conditions. Both close-talking and far-field data were included in this training set, hence it also included non-meeting data. 

\subsection{Multi-channel training data}
We used 165 sessions of real meeting data to train the front-end module. The meetings were recorded by using a 7-channel circular microphone array with one microphone in the center, as well as attendee-wise head-worn close-talking microphones. Hence, we were able to obtain two training sets with the same speech content - a real multi-channel set and a multi-channel set simulated from single-channel close-talking data. Note that the 60-hour close-talking segments were also included in 75K data for AED training. Information of these training sets is summarizezd in Table.\ref{tab:config}.

\subsubsection{Real meeting data}
\label{sec:realdata}
Considering the fact that even humans sometimes find it difficult to accurately transcribe overlapped speech in meeting recordings, the data (60 hours, denoted by \textbf{Real}) we used for training only contained non-overlap regions, which provided us with accurately segmented and labeled  multi-channel audio data. Furthermore, to support the purpose of overlapped speech extraction, we generated another 60-hour overlapped data (denoted by \textbf{Real+}) on top of the non-overlap one. Specifically, we classified the segments according to the speaker labels in each meeting session at first. Given each segment, we randomly chose an interference segment from another speaker in the same session, then got a random overlap ratio (1-100\%) and trimmed this segment based on the interference duration, calculated by the overlap ratio and length of the segment to be mixed. Finally, we simply added this trimmed multi-channel interference signal to the segment to be mixed. 

In this way, we manually generated another 60-hour overlapped data, while keeping the original transcription for each segment. The only assumption for this overlapped set is the voices' images from different speakers at the microphone array position are additive. It is much closer to what the real overlapped speech would be like than purely simulated multi-channel signals which we will describe in the following section.

\begin{table}[tb]
  \begin{center}
   	\caption{Training and evaluation data used for our meeting transcription experiments.}
    \label{tab:config}
\resizebox{0.95\linewidth}{!}{
	\begin{tabular}{ll}
	  \toprule
	  \toprule
	  {\scriptsize{}{\bf Training set}} \\
      {\scriptsize{}for Back-end} & {\scriptsize{} 75K hours, single-channel}\\
      {\scriptsize{}for Joint Model} & {\scriptsize{} 165 meeting sessions -} \\
                     & {\scriptsize{} \textbf{Simu}: 60 hours, 7-channel, simulation, non-overlapped} \\
                     & {\scriptsize{} \textbf{Real}: 60 hours, 7-channel, real, non-overlapped} \\
                     & {\scriptsize{} \textbf{Real+}: 60 hours, 7-channel, generated by mixing \textbf{Real}}\\
      \hline 
	 {\scriptsize{}{\bf Evaluation Set 1}} & {\scriptsize{}{\textbf{(Eval 1)}}} \\
	 {\scriptsize{}Num. of Sessions} & {\scriptsize{}9 real meetings}\\
	 {\scriptsize{}Overlap condition} & {\scriptsize{}28,690 / 8,041 words in non-overlapped / overlapped segments}\\
     {\scriptsize{}Speakers} & {\scriptsize{}4-17 attendees, unseen in training}\\
        \hline 
     {\scriptsize{}{\bf Evaluation Set 2}} & {\scriptsize{}{\textbf{(Eval 2)}}}\\
	 {\scriptsize{}Num. of Sessions} & {\scriptsize{}22 real meetings}\\
	 {\scriptsize{}Overlap condition} & {\scriptsize{}46,774 / 69,200 words in non-overlapped / overlapped segments}\\
     {\scriptsize{}Speakers} & {\scriptsize{}3-15 attendees, no restriction on the seen / unseen assumption}\\
      \bottomrule 
      \bottomrule
	\end{tabular}
}
  \end{center}
    \vspace{-0.6cm}
\end{table}

\subsubsection{Simulated meeting data}
Given the close-talking utterances with meeting session ids as well as the corresponding speaker label, we also created 60 hours of simulated multi-channel audio  (denoted by \textbf{Simu}) for training as with the previous studies~\cite{subramanian2020far, delcroix2020improving}. 
We tried to imitate the property of the multi-channel meeting recordings to the extent possible. Our simulation procedure was as follows,

\begin{enumerate}
    \item For each meeting session (165 sessions in total), we counted the number of speakers;
    \item We then randomly set the room size, length$\times$width$\times$height, within [4, 10]$m\times$[4, 10]$m\times$[2, 5]$m$ and the reverberation time, RT60, between 0.15$ms$ and 0.6$ms$;
    \item Assuming all the speakers did not move during the session, we randomly picked a point to put the 7-channel microphone array (with the same geometry as the real recording device) at a [1.0, 1.5]$m$ height. We also randomly determined the speakers' positions under the constraint that the speaker-to-array distance was 1.0$m$ or greater;
    \item The room impulse responses (RIR) between the speakers and the microphone array were derived with the image method \cite{allen1979image} and then applied to the corresponding speakers' close-talking segments;
    \item 7-channel diffuse noise signals were added to the reverberated signals with a signal-to-noise ratio (SNR) between -5$dB$ to 10$dB$. Finally, the simulated 7-channel signals were normalized to a certain volume.
\end{enumerate}

\subsection{Evaluation data}
We evaluated the proposed method on 
31 real meeting recordings, the statistics of which are presented in Table \ref{tab:config}.
According to the recording conditions,
they were divided into Eval 1 and Eval 2,
which contains the recordings of 9 sessions
and 22 sessions, respectively.
We used the reference start and end time of each utterance to extract evaluation segments, and 
each segment was categorized 
to either a non-overlapped or
overlapped segment based on the absence or presence of speech of interference speakers.
Note that we used the reference start and end time for our evaluation in order to evaluate the effectiveness of joint modeling without being affected by segmentation errors.
In actual applications, segmentation will be done by using speaker diaraztion \cite{yoshioka2019advances}.

The non-overlap/overlap distribution of Eval 1 is similar to the validation set we used to train the front-end module and the speakers in Eval 1 were unseen in the 60-hour training set. However, we did not have such speaker unseen restrictions for Eval 2 (more common in real applications), though the overlap ratio of which is higher than Eval 1.
Before decoding, we applied weighted prediction error (WPE) \cite{yoshioka2012generalization} to dereverbrate the multi-channel signals. 

\subsection{Model architecture and training procedure}
\label{sec:twostage}
Given that the size of the multi-channel training data is much smaller than that of the single-channel data, 
we applied a two-step training strategy.
In stage-1, we used the 75k-hour of single-channel data to train only the AED-based back-end; then in stage-2, the front-end module was appended to the network, and the parameters of the front-end module were updated by using the multi-channel data. 
The configurations of each module and training parameters are as follows. 

\subsubsection{Stage-1: Training AED-based back-end}
The AED consisted of 6 layers of 1024-dim bi-directional gated recurrent unit (GRU). Layer normalization was applied between the GRU layers. 
The decoder had 2 layers of 1024-dim uni-directional GRU. A conventional location-aware content-based attention with a single head was used \cite{chorowski2015attention}. 

The input feature for AED was 240-dimension log mel filterbanks, stacked by 3 frames with each frame having 10 msec. Global mean and variance normalization was applied to the features before feeding the features to the encoder. We used 32K mixed-unit with $\langle space\rangle$ symbol between words as recognition units \cite{li2018advancing}. Both teacher forcing \cite{chiu2018state} and label-smoothed cross-entropy loss \cite{chorowski2016towards} were applied in training. The batch size was 2500 frames and 64 GPUs were used. 
Both non-overlap and overlap as well as the overall segments were evaluated for each meeting session. Weighted average WERs for Eval 1 and Eval 2 were used as the performance metric to measure the effectiveness of each approach. Note that we didn't well fine-tune the training strategy for this back-end single-channel ASR model as we just used it as seed and baseline.


\begin{table}[tb]
  \begin{center}
   	\caption{Baseline and WER(\%) comparisons using simulated or real data for front-end training.}
   	\label{tab:baseline}
\resizebox{\linewidth}{!}{
	\begin{tabular}{cccccc}
\toprule
\toprule
      System & Training Data & Feature & Non-overlap & Overlap & Overall \\
\midrule
      {\it \hspace{-12mm}(Eval 1)}\\ 
      AED-ASR & 75K &  FBANK           & 17.79 & 34.37 & 21.42\\
\midrule
      E2E-MCASR & 75K\&\textbf{Simu} & 7-ch STFT              & 16.61 & 31.46  & 19.86 \\
      E2E-MCASR      & 75K\&\textbf{Real} & 7-ch STFT                   & \textbf{14.22} & 25.42 & \textbf{16.67} \\
      E2E-MCASR      & 75K\&\textbf{Real} & + IPD & 14.41               & \textbf{24.95} & 16.72 \\
\midrule
\midrule
      {\it\hspace{-12mm}(Eval 2)}\\ 
      AED-ASR & 75K & FBANK            & 15.59 & 30.40 & 24.42 \\
\midrule
      E2E-MCASR & 75K\&\textbf{Simu} & 7-ch STFT              & 14.96 & 28.64 & 23.12 \\
      E2E-MCASR  & 75K\&\textbf{Real} & 7-ch STFT                       & \textbf{13.99} & 25.23 & 20.70\\
      E2E-MCASR  & 75K\&\textbf{Real} & + IPD                           & 14.04 & \textbf{24.58} & \textbf{20.31} \\
\bottomrule
\bottomrule
	\end{tabular}
}
	\label{tab:two-stage}
  \end{center}
    \vspace*{-0.8cm}
\end{table}


\begin{table}[h]
  \begin{center}
   	\caption{WER (\%) of E2E MCASR with different proportions of overlapped segments in the training set. The training data is randomly sampled from \textbf{Real} (non-overlapped) and \textbf{Real+} (overlapped) under a constraint of the total data size. (NO: Non-overlap, OL: Overlap, All: Overall)} 
   	\label{tab:overlap}
\resizebox{\linewidth}{!}{
	\begin{tabular}{cccc}
	  \toprule
	  \toprule
      Total data & Ratio (\%) of &  \textbf{Eval 1} & \textbf{Eval 2} \\
      size (h)  & overlapped segments & NO / OL / All &  NO / OL / All \\
      \midrule
      60  & 0 & 14.41 / 24.95 / 16.72 & 14.04 / 24.58 / 20.31 \\
      60 & 30  & 14.36 / 23.45 / 16.35 & 14.17 / 23.90 / 19.98 \\
      60 & 60 & 14.44 / 22.59 / 16.22 & 14.17 / \textbf{23.76} / 19.90 \\
      60 & 100$^\dagger$ & \textbf{14.17} / \textbf{22.48} / \textbf{15.99} & 14.26 / 23.95 / 20.04 \\
\midrule
      120 & 50 & 14.33 / 22.65 / 16.16 & \textbf{14.05} / 23.85 / \textbf{19.90} \\
      \bottomrule
      \bottomrule
	\end{tabular}
}
	\\
    \begin{tablenotes}
    \footnotesize
   \item[*]$^\dagger$Note: the 100\% overlapped segments still contain non-overlapped regions, overlap ratio of each segment ranges from 1\% to 100\%.
    \end{tablenotes}
 \end{center}
    \vspace{-0.8cm}
\end{table}

\setlength{\dashlinedash}{2pt}
\setlength{\dashlinegap}{2pt}
\begin{table*}[t]
  \begin{center}
   	\caption{WERs(\%) of E2E MCASR using different multi-channel training data, models and bias information. (shallow/deep represents shallow/deep concatenation.)}
   	{\footnotesize
	\begin{tabular}{cccccc}
	  \toprule
	  \toprule
Training Data&  Location bias & Speaker bias & Session-based  &  \textbf{Eval 1}    & \textbf{Eval 2}              \\
 &  (Angle) &  (d-vector)   &  Bias Info.     & Non-overlap / Overlap / Overall  & Non-overlap / Overlap / Overall\\
      \midrule
 \textbf{Real} & -   & -    & -       & 14.41 / 24.95 / 16.72                   & 14.04 / 24.58 / 20.31 \\
 \textbf{Real} &  shallow & -               & - & {\bf 14.20} / 24.11 / 16.37                   & 14.04 / 23.91 / 19.93 \\ 
 \textbf{Real} &   shallow & -     & \checkmark & 14.23 / {\bf 23.61} / {\bf 16.28}    & 14.01 / {\bf 23.23} / {\bf 19.51} \\
 \textbf{Real} & -  & shallow  & -     & 14.38 / 24.03 / 16.49                     & {\bf 13.96} / 24.15 / 20.04  \\
 \textbf{Real} & - & shallow   & \checkmark & 14.41 / 24.08 / 16.53                & 14.00 / 24.05 / 20.00 \\
        \midrule
 \textbf{Real} \&  \textbf{Real+} &  - & -   & -       & 14.33 / 22.65 / 16.16                  & {\bf 14.05} / 23.85 / 19.90 \\
  \textbf{Real} \&  \textbf{Real+} &  shallow & - & - & 14.01 / 22.11 / 15.79                   & 14.24 / 23.61 / 19.83 \\
  \textbf{Real} \&  \textbf{Real+} &   shallow & -   & \checkmark & 14.06 / 20.85 / 15.57       & 14.22 / 21.29 / 18.44 \\
  \textbf{Real} \&  \textbf{Real+} &  deep & - & \checkmark & {\bf 13.80} / 20.69 / 15.31    & 14.06 / 20.92 / 18.15 \\
  \textbf{Real} \&  \textbf{Real+} &  deep & -  & + Pre-masking & 14.01 / {\bf 19.94} / {\bf 15.31}             & 14.19 / {\bf 20.24} / {\bf 17.79} \\ \hdashline
  \textbf{Real} \&  \textbf{Real+} &  deep & shallow$^{\dagger}$ & \checkmark &  14.13 / 19.88 / 15.38           & 14.09 / 20.61 / 17.98 \\
  \textbf{Real} \&  \textbf{Real+} &  deep & shallow$^{\dagger}$  & + Pre-masking & 14.25 / {\bf 19.85} / 15.47       & 14.33 / {\bf 19.96} / {\bf 17.69} \\
      \bottomrule
      \bottomrule
	\end{tabular}}
	\\
    \begin{tablenotes}
    \footnotesize
   \item[*] \hspace{15mm}$^\dagger$ We didn't apply deep concatenation for speaker bias because we observed degradation in our preliminary experiments.
    \end{tablenotes}
	\label{tab:biaseval}
  \end{center}
    \vspace{-0.5cm}
\end{table*}

\subsubsection{Stage-2: Jointly training the front-end}
The front-end module, specifically the mask estimation network, consisted of 2 BLSTM layers with each layer having 300 units. The front-end module was concatenated with the AED-based back-end via the feature transformation module which mapped the 257-dimension beamformer output to the 240-dimension log mel filterbanks. The input feature of the baseline front-end module was 257-dimension Short-time Fourier transform (STFT) of the multi-channel audios, concatenated with the IPD features which were calculated based on the microphone pairs $"(1, 0), (2, 0), (3, 0), (4, 0), \\
(5, 0), (6, 0)"$ - center microphone was used as the reference for the rest. The IPD features appended to the 7-channel STFT features were the same, which resulted in the feature dimension of each channel expanded from $257$ to $(1+6)*257$.

Given the back-end with parameters frozen that was trained in Stage-1, the front-end module of the joint model was trained using the Adam optimizer with a learning rate schedule similar to that described in \cite{park2019specaugment}. The learning rate was linearly increased from 0 to 0.0001 by using the initial 1k iterations, kept until the 120k-th iteration, then exponentially decaying to 0.00001 at 240k iterations. In this work, we report the results of validation best models found after 200k of training iterations. The minibatch was 3000 frames, and 8 V100 GPUs were used for all the trainings.

\subsection{Simulated v.s. real multi-channel training data}
Table \ref{tab:baseline} shows the WERs of the E2E MCASR models that were trained on simulated data or real data. We can make the following observations. 

\begin{itemize}
\setlength{\itemsep}{0pt}
\setlength{\parskip}{0pt}
    \item When the model was trained on the multi-channel STFT features derived from the simulated meeting data, the overall WERs for Eval 1 and Eval 2 got improved from 21.42\% to 19.86\% and from 24.42\% to 23.12\%, respectively. The improvement was observed but limited due to the mismatch between training and testing.
    \item When the real data were used for the training, the WERs were substantially improved to 16.67\% and 20.70\% for Eval 1 and Eval 2, respectively. 
    This implies that a substantial difference existed between the real and simulated training. It also suggests that the E2E MCASR could be sensitive to the mismatch of the training/testing conditions, and it is important to utilize the real training data even when the data quantity is limited.  
    \item Unbiased IPD features marginally improved the WER for overlapped regions while the WER for non-overlapped regions was very slightly degraded.
\end{itemize}

\subsection{Investigation of overlap proportion in training}
To further improve the model capability to handle overlapped segments, we introduced the overlapped data \textbf{Real+} for model training, where the overlapped data generation was described in Section \ref{sec:realdata}. Table \ref{tab:overlap} shows how the  use of the overlapped data during training affects the ASR performance, especially for the overlapped segments.

The baseline was the model trained on the original 60-hour multi-channel data. When we increased the ratio of overlapped segments to 30\% in the 60-hour training data, we observed that the WERs of overlapped segments got improved for both Eval 1 and Eval 2. With further increasing the ratio to 60\% or 100\%, the gain was marginal for Eval 2, but it was substantial for the overlapped segments of Eval 1. It could be because the non-overlap to overlap distribution of Eval 1 was more similar to that of the validation set.
Finally, when we simply joined \textbf{Real} and \textbf{Real+} to form a 120-hour training set, we obtained overall WERs of 16.16\% for Eval 1 and 19.90\% for Eval 2.
These results were close to 60-hour training with the data having 100\% overlapped segments. 
It suggests that simply duplicating the real data with the same content would not benefit the E2E MCASR front-end training. Due to that the performances of 120-hour training and 60-hour fully (100\%) overlapped training were on par, we just used the 120-hour set to train the model with biased information in the following experiments, representing that we have included overlapped segments in training.

\subsection{Evaluation with bias information}
Table \ref{tab:biaseval} shows the WERs of the E2E MCASR models with different training sets and different kinds of bias information. Compared to the model that was trained on the 60-hour non-overlap data (\textbf{Real}) using unbiased features, shallowly concatenating the angle feature or d-vector to the model could improve the ASR performance for both Eval 1 and Eval 2. Greater WER reduction was observed for the overlapped segments.
When session-based decoding was applied, the model with the location bias achieved additional gains thanks to more accurate DOA estimation. Meanwhile, using a session-based d-vector for each speaker did not show any superiority, which means the d-vector is not sensitive to partially overlapped segments.
Overall, it suggests that both location and speaker biases are helpful for the E2E MCASR, especially for overlapped segments. The gain from the location bias is larger than that from the speaker bias as all directions can be covered in training.

Given the fact that the location bias yielded greater WER reduction, we continued investigation by using the angle-feature-based joint model trained on \textbf{Real}\&\textbf{Real+} as the basis. 
The overall WERs obtained by using the location bias for Eval 1 and Eval 2 were 15.57\% and 18.44\%, respectively, and were improved from 16.16\% and 19.90\% obtained with the unbiased features. 
The deep concatenation of the location bias further reduced the WERs for both sets (15.31\% and 18.15\%). 
Using the shallow-concatenated d-vector on top of the deep concatenation of the angle feature yielded better WERs for overlapped segments, as can be seen in the last two rows of Table \ref{tab:biaseval}.
The extra pre-masking process greatly improved the WERs for the overlapped segments, while it slightly degraded the WERs for the non-overlapped segments. 
Applying pre-masking benefited more for Eval 2 which had more overlapped segments.
We finally obtained the WERs of 15.47\% for Eval 1 and 17.69\% for Eval 2 with the training and decoding strategies, suggested by Fig.\ref{fig:dftnet} and the last row of Table \ref{tab:biaseval}.

\section{Conclusion}
This work investigated application of E2E MCASR to offline meeting transcription under practical settings. 
We started from an AED-based baseline ASR model trained with 75k hours of single-channel data, then improved the ASR performance of both non-overlapped and overlapped segments. 
We presented a practical way for leveraging a relatively limited amount of multi-channel data for joint front-end/back-end training and examined the impact of using real multi-channel training data.
In addition, the ASR performance for overlapped segments was further improved by introducing the bias information which helps the front-end module focus on target speakers.
Our meeting transcription experiment results showed that the framework could largely benefit from the real data training and the use of the bias information.
10+\% relative WER reduction was observed by replacing the simulation data with the real one.
Overall, we achieved 27\% relative WER reduction on real meeting recordings compared with the strong single-channel model trained on the large quantity of data. 


\bibliographystyle{IEEEbib}
\bibliography{refs}

\begin{thebibliography}{10}

\bibitem{xiong2016achieving}
Wayne Xiong, Jasha Droppo, Xuedong Huang, Frank Seide, Mike Seltzer, Andreas
  Stolcke, Dong Yu, and Geoffrey Zweig,
\newblock ``Achieving human parity in conversational speech recognition,''
\newblock {\em arXiv preprint arXiv:1610.05256}, 2016.

\bibitem{yoshioka2019advances}
Takuya Yoshioka, Igor Abramovski, Cem Aksoylar, Zhuo Chen, Moshe David,
  Dimitrios Dimitriadis, Yifan Gong, Ilya Gurvich, Xuedong Huang, Yan Huang,
  et~al.,
\newblock ``Advances in online audio-visual meeting transcription,''
\newblock in {\em 2019 IEEE Automatic Speech Recognition and Understanding
  Workshop (ASRU)}. IEEE, 2019, pp. 276--283.

\bibitem{chen2020continuous}
Zhuo Chen, Takuya Yoshioka, Liang Lu, Tianyan Zhou, Zhong Meng, Yi~Luo, Jian
  Wu, Xiong Xiao, and Jinyu Li,
\newblock ``Continuous speech separation: Dataset and analysis,''
\newblock in {\em ICASSP 2020-2020 IEEE International Conference on Acoustics,
  Speech and Signal Processing (ICASSP)}. IEEE, 2020, pp. 7284--7288.

\bibitem{li2017acoustic}
Bo~Li, Tara~N Sainath, Arun Narayanan, Joe Caroselli, Michiel Bacchiani, Ananya
  Misra, Izhak Shafran, Hasim Sak, Golan Pundak, Kean~K Chin, et~al.,
\newblock ``Acoustic modeling for google home.,''
\newblock in {\em Interspeech}, 2017, pp. 399--403.

\bibitem{yoshioka2018recognizing}
Takuya Yoshioka, Hakan Erdogan, Zhuo Chen, Xiong Xiao, and Fil Alleva,
\newblock ``Recognizing overlapped speech in meetings: A multichannel
  separation approach using neural networks,''
\newblock {\em Proc. Interspeech 2018}, pp. 3038--3042, 2018.

\bibitem{yoshioka2018multi}
Takuya Yoshioka, Hakan Erdogan, Zhuo Chen, and Fil Alleva,
\newblock ``Multi-microphone neural speech separation for far-field
  multi-talker speech recognition,''
\newblock in {\em 2018 IEEE International Conference on Acoustics, Speech and
  Signal Processing (ICASSP)}. IEEE, 2018, pp. 5739--5743.

\bibitem{wang2018supervised}
DeLiang Wang and Jitong Chen,
\newblock ``Supervised speech separation based on deep learning: An overview,''
\newblock {\em IEEE/ACM Transactions on Audio, Speech, and Language
  Processing}, vol. 26, no. 10, pp. 1702--1726, 2018.

\bibitem{kinoshita2017neural}
Keisuke Kinoshita, Marc Delcroix, Haeyong Kwon, Takuma Mori, and Tomohiro
  Nakatani,
\newblock ``Neural network-based spectrum estimation for online wpe
  dereverberation,''
\newblock {\em Proc. Interspeech 2017}, pp. 384--388, 2017.

\bibitem{heymann2016neural}
J.~Heymann, L.~Drude, and R.~Haeb-Umbach,
\newblock ``Neural network based spectral mask estimation for acoustic
  beamforming,''
\newblock in {\em ICASSP}, 2016, pp. 196--200.

\bibitem{erdogan2016improved}
Hakan Erdogan, John~R Hershey, Shinji Watanabe, Michael~I Mandel, and Jonathan
  Le~Roux,
\newblock ``Improved mvdr beamforming using single-channel mask prediction
  networks.,''
\newblock in {\em Interspeech}, 2016, pp. 1981--1985.

\bibitem{yu2017permutation}
Dong Yu, Morten Kolb{\ae}k, Zheng-Hua Tan, and Jesper Jensen,
\newblock ``Permutation invariant training of deep models for
  speaker-independent multi-talker speech separation,''
\newblock in {\em 2017 IEEE International Conference on Acoustics, Speech and
  Signal Processing (ICASSP)}. IEEE, 2017, pp. 241--245.

\bibitem{boeddeker2018front}
Christoph Boeddeker, Jens Heitkaemper, Joerg Schmalenstroeer, Lukas Drude, Jahn
  Heymann, and Reinhold Haeb-Umbach,
\newblock ``Front-end processing for the chime-5 dinner party scenario,''
\newblock in {\em CHiME5 Workshop, Hyderabad, India}, 2018.

\bibitem{carletta2005ami}
Jean Carletta, Simone Ashby, Sebastien Bourban, Mike Flynn, Mael Guillemot,
  Thomas Hain, Jaroslav Kadlec, Vasilis Karaiskos, Wessel Kraaij, Melissa
  Kronenthal, et~al.,
\newblock ``The ami meeting corpus: A pre-announcement,''
\newblock in {\em International workshop on machine learning for multimodal
  interaction}. Springer, 2005, pp. 28--39.

\bibitem{barker2018fifth}
Jon Barker, Shinji Watanabe, Emmanuel Vincent, and Jan Trmal,
\newblock ``The fifth'chime'speech separation and recognition challenge:
  dataset, task and baselines,''
\newblock {\em arXiv preprint arXiv:1803.10609}, 2018.

\bibitem{watanabe2020chime}
Shinji Watanabe, Michael Mandel, Jon Barker, and Emmanuel Vincent,
\newblock ``Chime-6 challenge: Tackling multispeaker speech recognition for
  unsegmented recordings,''
\newblock {\em arXiv preprint arXiv:2004.09249}, 2020.

\bibitem{erdogan2016multi}
H.~Erdogan, T.~Hayashi, J.~R. Hershey, et~al.,
\newblock ``Multi-channel speech recognition: Lstms all the way through,''
\newblock in {\em CHiME-4 workshop}, 2016, pp. 1--4.

\bibitem{kanda2018hitachi}
Naoyuki Kanda, Rintaro Ikeshita, Shota Horiguchi, Yusuke Fujita, Kenji
  Nagamatsu, Xiaofei Wang, Vimal Manohar, Nelson Enrique~Yalta Soplin, Matthew
  Maciejewski, Szu-Jui Chen, et~al.,
\newblock ``The {Hitachi/JHU CHiME-5} system: Advances in speech recognition
  for everyday home environments using multiple microphone arrays,''
\newblock in {\em Proc. CHiME-5}, 2018, pp. 6--10.

\bibitem{kanda2019acoustic}
Naoyuki Kanda, Yusuke Fujita, Shota Horiguchi, Rintaro Ikeshita, Kenji
  Nagamatsu, and Shinji Watanabe,
\newblock ``Acoustic modeling for distant multi-talker speech recognition with
  single-and multi-channel branches,''
\newblock in {\em Proc. ICASSP}, 2019, pp. 6630--6634.

\bibitem{subramanian2019speech}
Aswin~Shanmugam Subramanian, Xiaofei Wang, Murali~Karthick Baskar, Shinji
  Watanabe, Toru Taniguchi, Dung Tran, and Yuya Fujita,
\newblock ``Speech enhancement using end-to-end speech recognition
  objectives,''
\newblock in {\em 2019 IEEE Workshop on Applications of Signal Processing to
  Audio and Acoustics (WASPAA)}. IEEE, 2019, pp. 234--238.

\bibitem{subramanian2020far}
Aswin~Shanmugam Subramanian, Chao Weng, Meng Yu, Shi-Xiong Zhang, Yong Xu,
  Shinji Watanabe, and Dong Yu,
\newblock ``Far-field location guided target speech extraction using end-to-end
  speech recognition objectives,''
\newblock in {\em ICASSP 2020-2020 IEEE International Conference on Acoustics,
  Speech and Signal Processing (ICASSP)}. IEEE, 2020, pp. 7299--7303.

\bibitem{von2020multi}
Thilo von Neumann, Christoph Boeddeker, Lukas Drude, Keisuke Kinoshita, Marc
  Delcroix, Tomohiro Nakatani, and Reinhold Haeb-Umbach,
\newblock ``Multi-talker asr for an unknown number of sources: Joint training
  of source counting, separation and asr,''
\newblock {\em arXiv preprint arXiv:2006.02786}, 2020.

\bibitem{zhang2020end}
Wangyou Zhang, Aswin~Shanmugam Subramanian, Xuankai Chang, Shinji Watanabe, and
  Yanmin Qian,
\newblock ``End-to-end far-field speech recognition with unified
  dereverberation and beamforming,''
\newblock {\em arXiv preprint arXiv:2005.10479}, 2020.

\bibitem{sainath2017multichannel}
Tara~N Sainath, Ron~J Weiss, Kevin~W Wilson, Bo~Li, Arun Narayanan, Ehsan
  Variani, Michiel Bacchiani, Izhak Shafran, Andrew Senior, Kean Chin, et~al.,
\newblock ``Multichannel signal processing with deep neural networks for
  automatic speech recognition,''
\newblock {\em IEEE/ACM Transactions on Audio, Speech, and Language
  Processing}, vol. 25, no. 5, pp. 965--979, 2017.

\bibitem{meng2017deep}
Zhong Meng, Shinji Watanabe, John~R Hershey, and Hakan Erdogan,
\newblock ``Deep long short-term memory adaptive beamforming networks for
  multichannel robust speech recognition,''
\newblock in {\em 2017 IEEE International Conference on Acoustics, Speech and
  Signal Processing (ICASSP)}. IEEE, 2017, pp. 271--275.

\bibitem{minhua2019frequency}
Wu~Minhua, Kenichi Kumatani, Shiva Sundaram, Nikko Str{\"o}m, and Bj{\"o}rn
  Hoffmeister,
\newblock ``Frequency domain multi-channel acoustic modeling for distant speech
  recognition,''
\newblock in {\em ICASSP 2019-2019 IEEE International Conference on Acoustics,
  Speech and Signal Processing (ICASSP)}. IEEE, 2019, pp. 6640--6644.

\bibitem{chen2018multi}
Zhuo Chen, Xiong Xiao, Takuya Yoshioka, Hakan Erdogan, Jinyu Li, and Yifan
  Gong,
\newblock ``Multi-channel overlapped speech recognition with location guided
  speech extraction network,''
\newblock in {\em 2018 IEEE Spoken Language Technology Workshop (SLT)}. IEEE,
  2018, pp. 558--565.

\bibitem{delcroix2020improving}
Marc Delcroix, Tsubasa Ochiai, Katerina Zmolikova, Keisuke Kinoshita, Naohiro
  Tawara, Tomohiro Nakatani, and Shoko Araki,
\newblock ``Improving speaker discrimination of target speech extraction with
  time-domain speakerbeam,''
\newblock in {\em ICASSP 2020-2020 IEEE International Conference on Acoustics,
  Speech and Signal Processing (ICASSP)}. IEEE, 2020, pp. 691--695.

\bibitem{xiao2019single}
Xiong Xiao, Zhuo Chen, Takuya Yoshioka, Hakan Erdogan, Changliang Liu,
  Dimitrios Dimitriadis, Jasha Droppo, and Yifan Gong,
\newblock ``Single-channel speech extraction using speaker inventory and
  attention network,''
\newblock in {\em ICASSP 2019-2019 IEEE International Conference on Acoustics,
  Speech and Signal Processing (ICASSP)}. IEEE, 2019, pp. 86--90.

\bibitem{ochiai2017unified}
Tsubasa Ochiai, Shinji Watanabe, Takaaki Hori, John~R Hershey, and Xiong Xiao,
\newblock ``Unified architecture for multichannel end-to-end speech recognition
  with neural beamforming,''
\newblock {\em IEEE Journal of Selected Topics in Signal Processing}, vol. 11,
  no. 8, pp. 1274--1288, 2017.

\bibitem{delcroix2019end}
Marc Delcroix, Shinji Watanabe, Tsubasa Ochiai, Keisuke Kinoshita, Shigeki
  Karita, Atsunori Ogawa, and Tomohiro Nakatani,
\newblock ``End-to-end speakerbeam for single channel target speech
  recognition.,''
\newblock 2019.

\bibitem{benesty2008microphone}
Jacob Benesty, Jingdong Chen, and Yiteng Huang,
\newblock {\em Microphone array signal processing}, vol.~1,
\newblock Springer Science \& Business Media, 2008.

\bibitem{zhang2007maximum}
Cha Zhang, Zhengyou Zhang, and Dinei Flor{\^e}ncio,
\newblock ``Maximum likelihood sound source localization for multiple
  directional microphones,''
\newblock in {\em 2007 IEEE International Conference on Acoustics, Speech and
  Signal Processing-ICASSP'07}. IEEE, 2007, vol.~1, pp. I--125.

\bibitem{nagrani2017voxceleb}
Arsha Nagrani, Joon~Son Chung, and Andrew Zisserman,
\newblock ``Voxceleb: A large-scale speaker identification dataset,''
\newblock in {\em Proc. Interspeech}, 2017, pp. 2616--2620.

\bibitem{chung2018voxceleb2}
Joon~Son Chung, Arsha Nagrani, and Andrew Zisserman,
\newblock ``Voxceleb2: Deep speaker recognition,''
\newblock in {\em Proc. Interspeech}, 2018, pp. 1086--1090.

\bibitem{zhou2019cnn}
Tianyan Zhou, Yong Zhao, Jinyu Li, Yifan Gong, and Jian Wu,
\newblock ``Cnn with phonetic attention for text-independent speaker
  verification,''
\newblock in {\em 2019 IEEE Automatic Speech Recognition and Understanding
  Workshop (ASRU)}. IEEE, 2019, pp. 718--725.

\bibitem{allen1979image}
Jont~B Allen and David~A Berkley,
\newblock ``Image method for efficiently simulating small-room acoustics,''
\newblock {\em The Journal of the Acoustical Society of America}, vol. 65, no.
  4, pp. 943--950, 1979.

\bibitem{yoshioka2012generalization}
Takuya Yoshioka and Tomohiro Nakatani,
\newblock ``Generalization of multi-channel linear prediction methods for blind
  mimo impulse response shortening,''
\newblock {\em IEEE Transactions on Audio, Speech, and Language Processing},
  vol. 20, no. 10, pp. 2707--2720, 2012.

\bibitem{chorowski2015attention}
Jan~K Chorowski, Dzmitry Bahdanau, Dmitriy Serdyuk, Kyunghyun Cho, and Yoshua
  Bengio,
\newblock ``Attention-based models for speech recognition,''
\newblock in {\em Advances in neural information processing systems}, 2015, pp.
  577--585.

\bibitem{li2018advancing}
Jinyu Li, Guoli Ye, Amit Das, Rui Zhao, and Yifan Gong,
\newblock ``Advancing acoustic-to-word ctc model,''
\newblock in {\em 2018 IEEE International Conference on Acoustics, Speech and
  Signal Processing (ICASSP)}. IEEE, 2018, pp. 5794--5798.

\bibitem{chiu2018state}
Chung-Cheng Chiu, Tara~N Sainath, Yonghui Wu, Rohit Prabhavalkar, Patrick
  Nguyen, Zhifeng Chen, Anjuli Kannan, Ron~J Weiss, Kanishka Rao, Ekaterina
  Gonina, et~al.,
\newblock ``State-of-the-art speech recognition with sequence-to-sequence
  models,''
\newblock in {\em 2018 IEEE International Conference on Acoustics, Speech and
  Signal Processing (ICASSP)}. IEEE, 2018, pp. 4774--4778.

\bibitem{chorowski2016towards}
Jan Chorowski and Navdeep Jaitly,
\newblock ``Towards better decoding and language model integration in sequence
  to sequence models,''
\newblock {\em arXiv preprint arXiv:1612.02695}, 2016.

\bibitem{park2019specaugment}
Daniel~S Park, William Chan, Yu~Zhang, Chung-Cheng Chiu, Barret Zoph, Ekin~D
  Cubuk, and Quoc~V Le,
\newblock ``Specaugment: A simple data augmentation method for automatic speech
  recognition,''
\newblock in {\em Proc. Interspeech}, 2019, pp. 2613--2617.

\end{thebibliography}

\end{document}